\documentclass[preprint]{aastex}

\received{}
\accepted{}
\journalid{}{}
\articleid{}{}

\shorttitle{Microlensing and the mass function of halo dark matter}
\shortauthors{Green}

\begin{document}

\title{Probing the mass function of halo dark matter via microlensing}

\author{Anne M. Green}
\affil{Astronomy Unit, School of Mathematical Sciences, QMW, Mile End Road, 
London, \\ E1 4NS, UK}
\altaffiltext{1}{amg@maths.qmw.ac.uk}

\begin{abstract}
The simplest interpretation of the microlensing events observed
towards the Large Magellanic Clouds is that approximately half of the
mass of the Milky Way halo is in the form of MAssive Compact Halo
Objects with $M \sim 0.5 M_{\odot}$. It is not possible, due to limits
from star counts and chemical abundance arguments, for faint stars or
white dwarves to comprise such a large fraction of the halo mass. This
leads to the consideration of more exotic lens candidates, such as
primordial black holes, or alternative lens locations. If the lenses
are located in the halo of the Milky Way, then constraining their mass
function will shed light on their nature. Using the current
microlensing data we find, for four halo models, the best fit
parameters for delta-function, primordial black hole and various power
law mass functions. The best fit primordial black hole mass functions,
despite having significant finite width, have likelihoods which are
similar to, and for one particular halo model greater than, those of
the best fit delta functions .  We then use Monte Carlo simulations to
investigate the number of microlensing events necessary to determine
whether the MACHO mass function has significant finite width. If the
correct halo model is known, then $\sim$ 500 microlensing events will
be sufficient, and will also allow determination of the mass function
parameters to $\sim 5\%$.
\end{abstract}


\section{Introduction}
The rotation curves of spiral galaxies are typically flat out to about
$\sim 30$ kpc. This implies that the mass enclosed increases linearly
with radius, with a halo of dark matter extending beyond the luminous
matter (Ashman 1992; Freeman 1995; Kochanek 1995). The nature of the
dark matter is unknown (see e.g. Primack, Sadoulet \& Seckel 1988),
with possible candidates including massive astrophysical compact
objects (MACHOs), such as brown dwarves, Jupiters or black holes and
elementary particles, known as Weakly Interacting Massive Particles
(WIMPs), such as axions and neutrilinos.

MACHOs with mass in the range $10^{-8} M_{\odot}$ to $10^{3}
M_{\odot}$ can be detected via the temporary amplification of
background stars which occurs, due to gravitational microlensing, when
the MACHO passes close to the line of sight to a background star
(Paczy\'{n}ski 1986). Since the early 1990s several collaborations
have been monitoring millions of stars in the Large and Small
Magellanic Clouds (LMC and SMC), and a number of candidate
microlensing events have been observed.

The interpretation of these microlensing events is a matter of much
debate. Whilst the lenses responsible for these events could be
located in the halo of our galaxy, it is possible that the
contribution to the lensing rate due to other populations of objects
has been underestimated (for a discussion see e.g. Bennett 1998; Zhao
1999).

In the case of the 2 events observed towards the SMC there is
significant evidence that, in both cases, the lenses are in fact
located within the SMC. The first of these events (97-SMC-1) had a
duration of around 217 days, which implies that if the lens is in the
Milky Way halo then it probably has a mass in excess of $2 M_{\odot}$
(Palanque-Delabrouille et. al. 1997; Alcock et.al. 1997b; Sahu \& Sahu
1999). Spectroscopy of the source does not show the contamination
which would be expected if the lens was a star with such a high mass
(Sahu \& Sahu 1999). This leaves two possibilities: either the lens is
a low mass star in the SMC, or the lens is in the Milky Way halo but
is non-stellar. The second event (98-SMC-1) was a binary, which
allowed the time taken for the lens to cross the source star, and
hence the projected velocity of the lens, to be measured (Afonso
et. al. 1998; Alcock et. al. 1999). The probability of a standard halo
lens having a projected velocity as low as that measured ($v_{{\rm
proj}}=84$ km ${\rm s}^{-1}$) is of order $0.2 \%$ (Alcock
et. al. 1999). Due to tidal disruption, however, the SMC is elongated
along the line of sight, and hence its self-lensing rate can be high
enough to account for both the observed events (see Gyuk, Dalal, \&
Griest (1999) and references therein).

 Of the 8 events observed towards the LMC by the MACHO collaboration
during their first two years of observations, one event was a due to a
binary lens with a very low lens projected velocity, $v_{{\rm
proj}}=19$ km ${\rm s}^{-1}$. Alcock et. al. (1996a) argue that if the
LMC self-lensing optical depth is large enough to be consistent with
all 8 lenses being located in the LMC disk, then the probability of
finding a projected velocity value this low is less than
$0.5$\%. Therefore, whilst this lens may itself be in the LMC disk it
is unlikely that all 8 events are due to lenses in the LMC disk. If
the source star is itself also a binary, however, then the low
projected velocity is consistent with the lens being located in either
the Milky Way halo or the LMC disk (Alcock et. al. 1996a). It has also
been argued that there is selection bias against the observation of
halo binaries (Honma 1999).  These arguments do not, however, rule out
self-lensing by other LMC populations. Models of the LMC which have a
self-lensing optical depth large enough to account for the observed
events have been constructed (Aubourg et. al. 1999; Salati
et. al. 1999; Evans \& Kerins 1999). These models require the LMC
lenses to be distributed in an extended, shroud or halo like
distribution. It was previously though that, if the lenses are
stellar, there were difficulties reconciling such an extended
distribution with the low velocity dispersions observed ( see i.e Gyuk
et. al. 1999). Recent analysis of the radial velocities of Carbon
stars by Graff et. al. (1999), however, provides evidence for multiple
stellar components. The LMC could also have a dark matter halo with
the same MACHO mass fraction as the Milky Way, which would make a
significant contribution to the microlensing optical depth (Kerins \&
Evans 1998).  Other possible locations for the lenses include a dark
galaxy, or tidal debris, along the line of sight to the LMC (Zaritsky
\& Lin 1997; Zhao 1998; Zaritsky et. al. 1999), a warped and flared MW
disk (Evans et. al. 1998) and an extended MW protodisk (Gyuk \& Gates
1999).

There are several long-term prospects for unambiguously determining
the location of the lenses responsible for the LMC events. These
include a more sensitive micro-lensing survey, covering the whole of
the LMC, such as the proposed `SuperMACHO survey' (Stubbs 1998),
parallax observations, by a satellite such as the Space Interformetry
Mission, (Boden, Shao, \& Van Buren 1998; Gould \& Salim 1999) and
microlensing searches towards M31 (Ansari et. al. 1997; Gyuk \& Crotts
1999).  In the meantime the location of the lenses is an open
question.  In this paper we will subsequently assume that the events
observed towards the LMC are caused by MACHOs located in the halo of
our galaxy.

Since the duration of a microlensing event depends on the position,
transverse velocity and mass of the lens it is not possible to
associate a unique MACHO mass with each event. However there are two
techniques which can be used, assuming a specific halo model, to probe
the mass function of the MACHOs: maximum likelihood fitting of a
parametrised mass function (Alcock et.al. 1996b, 1997a; Mao \&
Paczy\'{n}ski 1996) and the method of mass moments (De Rujula
et. al. 1991; Jetzer 1994, Mao \& Paczy\'{n}ski 1996). Mao and
Paczy\'{n}ski (1996) found that the maximum likelihood method, whilst
slower than the mass moment method, is more robust. For the standard
halo model, a cored isothermal sphere, the most likely MACHO mass
function is sharply peaked around $0.5 M_{\odot}$, with about half of
the total mass of the halo in MACHOs (Alcock et. al. 1997a). This
poses a problem for stellar MACHO candidates. In the case of white
dwarves, an unreasonably large fraction of the baryons in the universe
would have had to have been cycled through the MACHOs and their
progenitors (see Freese, Fields \& Graff 1999 and references
therein). Direct searches place tight limits on the halo fraction in
faint stars (Charlot \& Silk 1995), however recent calculations have
found that old white dwarves with hydrogen dominated atmospheres may
in fact be blue (Hansen 1999a, 1999b), rather than red as previously
thought. The Hubble Deep Field South contains a number ( $\sim 5$) of
unresolved blue objects with high proper motions which may be old
white dwarves (Ibata et. al 1999). It is has also been argued however
that these objects are more likely to be planetary nebulae (Johnson
et. al. 1999) and furthermore no similar objects have been found in
ground--based proper motion studies (Flynn et. al. 1999).

These problems lead to the consideration of more exotic MACHO
candidates such as primordial black holes (PBHs) (Carr 1994).  PBHs
can be formed in the early universe, via a number of mechanisms, the
simplest of which is the collapse of large density perturbations
produced by inflation. In particular PBHs with mass $M \sim 0.5
M_{\odot}$ could be formed due to a spike in the primordial density
perturbation spectrum at this scale (Ivanov, Naselsky \& Novikov 1994;
Yokoyama 1995; Randall, Solja\u{c}i\'{c} \& Guth 1996;
Garc\'{\i}a-Bellido, Linde \& Wands 1996) or at the QCD phase
transition (Crawford \& Schramm 1982; Jedamzik 1997; Schwarz, Schmid
\& Widerin 1997; Schmid, Schwarz \& Widerin 1999; Jedamzik \& Niemeyer
1999) where the reduced pressure forces allow PBHs to form more
easily. In both cases it is not possible to produce an arbitrarily
narrow PBH mass function (Niemeyer \& Jedamzik 1998, 1999) and the
predicted mass function is considerably wider (Green \& Liddle 1999),
than the sharply peaked mass functions which have been fitted to the
observed events to date (Alcock 1997a).

Given the difficulties with stellar MACHO candidates it is therefore
important to investigate whether more exotic MACHO candidates, such as
PBHs, are compatible with the microlensing events observed towards the
LMC. In this paper we first compare the likelihood of the
delta--function and power--law mass functions, previously fitted to
the durations of the observed microlensing events, with that of the
PBH mass function. As well as the standard halo model we use Evans'
power--law halo models (Evans 1993, 1994; Alcock et. al. 1995, 1996b)
and also investigate the effect of incorporating the transverse
velocities of the source and observer (Griest 1991).  We then use
Monte Carlo simulations to address the question of the number of
events necessary to determine whether the MACHO mass function has
significant finite width.

\section{Primordial black hole formation and mass function}
\subsection{Collapse of density perturbations during radiation domination}
\label{defpbh}
In the early universe, where radiation dominates the equation of
state, for a PBH to form a collapsing region must be overdense enough
to overcome the pressure force resisting its collapse, as it falls
within its Schwarzschild radius. This occurs if the size of the
perturbation, $\delta=\delta \rho/ \rho$, is bigger than a critical
size, $\delta_{{\rm c}}$, at the time at which it enters the
horizon. There is also an upper limit of $\delta < 1$, since a
perturbation which exceeded this value would form a separate closed
universe (Harrison 1970). Early analytic calculations (Carr \& Hawking
1974) found $\delta_{{\rm c}} \sim 1/3$ with all PBHs having mass
roughly equal to the mass within the horizon at that time, known as
the horizon mass $M_{{\rm H}}$, independent of the size of the
perturbation.  Recent studies (Niemeyer \& Jedamzik 1998, 1999) of the
evolution of density perturbations have found that the mass of the PBH
formed in fact depends on the size of the perturbation:
\begin{equation}
\label{mbh}
M_{{\rm BH}}= k M_{{\rm H}} (\delta-\delta_{{\rm c}})^{\gamma} \,,
\end{equation}
where $\gamma \approx 0.37$ and $k$ and $\delta_{{\rm c}}$ are
constant for a given perturbation shape (for Mexican Hat shaped
fluctuations $k=2.85$ and $\delta_{{\rm c}}=0.67$), and 
\begin{equation}
\label{mhordef}
M_{\rm{H}} \approx \frac{4 \pi}{3} \rho (H^{-1})^{3} \,,
\end{equation}
where $\rho$ is the energy density and $H$ is the Hubble parameter.
In order to determine the number of PBHs formed on a given scale, and
hence the PBH mass function, we must smooth the density distribution
using a window function, $W(kR)$ (see e.g. Green \& Liddle 1997). For
Gaussian distributed fluctuations the probability distribution of the
smoothed density field $p(\delta(M_{{\rm H}}))$ is given by
\begin{equation}
\label{pdel}
p(\delta(M_{{\rm H}})) \, {\rm d} \delta(M_{{\rm H}}) = \frac{1}{ 
         \sqrt{2 \pi} \sigma(M_{{\rm H}})} 
	 \exp{\left[ - \frac{\delta^2(M_{{\rm H}})}
        {2 \sigma^2(M_{{\rm H}})}\right]} \, 
	{\rm d} \delta(M_{{\rm H}}) \,,
\end{equation}
where $\sigma(M_{{\rm H}})$ is the mass variance evaluated at horizon
crossing defined as in Liddle \& Lyth (1993)
\begin{equation}
\label{sigmam}
\sigma^2(M) = \frac{1}{2 \pi^2} \int_{0}^{\infty} P(k) W^2(kR) k^2 
	\,{\rm d}k \,,
\end{equation}
where $P(k) = \langle |\delta_{{\bf k}} |^2 \rangle$ is the power
spectrum.

The formation of PBHs on a range of scales has recently been studied
(Green \& Liddle 1999), for both power-law power spectra and flat
spectra with a spike on a given scale. In both cases it was found
that, in the limit where the number of PBHs formed is small enough to
satisfy the observational constraints on their abundance at
evaporation and at the present day, it can be assumed that all the
PBHs form at a single horizon mass.  It is therefore possible to
calculate the PBH mass distribution analytically:
\begin{equation}
\psi(M_{{\rm BH}}) = \frac{M_{{\rm BH}}}{M_{{\rm H}}} 
                  p(\delta(M_{{\rm H}})) 
                 \frac{{\rm d} \delta}{{\rm d} M_{{\rm BH}}} \,,
\end{equation}
which using eqs.~(\ref{mbh}) and (\ref{pdel}) becomes
\begin{equation}
\label{PBHmf}
\psi(M_{{\rm BH}})=\frac{1}{\sqrt{2 \pi} \gamma \sigma 
                  M_{{\rm H}}} \left( \frac{M_{{\rm BH}}}{k M_{{\rm H}}}
                   \right)^{\gamma}  
                \exp{ \left\{ - \frac{\left[\delta_{{\rm c}} +  
                 \left( \frac{M_{{\rm BH}}}{k M_{{\rm H}}}
                   \right)^{\gamma} \right]^2}{2 \sigma^2} \right\} } \,.
\end{equation}
The fraction of the total energy density in the universe, $\rho_{{\rm tot}}$,
 in the form
of PBHs, at the time they form, denoted by `$\rm{i}$', is then given by
\begin{equation}
\beta_{{\rm i}}= \frac{ \rho_{{\rm BH}}}{\rho_{{\rm tot}}} 
          = \int_{0}^{M_{{\rm max}}}
          \psi(M_{{\rm BH}})\, {\rm d} M_{{\rm BH}} \,,
\end{equation}
where $M_{{\rm max}}$ is the mass of the largest PBH which can form at
any given $M_{{\rm H}}$:
\begin{equation}
M_{{\rm max}}= k M_{{\rm H}} (1- \delta_{{\rm c}})^{\gamma} \,.
\end{equation}

The energy density in radiation dilutes as $\rho_{\rm{rad}} \propto
a^{-4}$, where $a$ is the scale factor, whereas that in PBHs decreases
more slowly, $\rho_{\rm{pbh}} \propto a^{-3}$ so that during radiation
domination the fraction of the energy density of the universe in PBHs
increases with time:
\begin{equation}
\beta \propto a \propto t^{1/2} \propto M_{{\rm H}}^{1/2}
\end{equation}
The present day abundance of PBHs must not exceed the maximum value
set by the present age and expansion rate of the universe (Carr 1975):
\begin{equation}
\label{opbh}
\Omega_{{\rm BH,0}}  = \Omega_{\rm{BH,eq}} <1 \,,
\end{equation}
where `eq' denotes the epoch of matter--radiation equality, after
which the density of PBHs, relative to the critical density, remains
constant. This constraint can be evolved backwards in time to
constrain $\beta_{{\rm i}}$, and hence $\sigma$:
\begin{equation}
\Omega_{\rm{BH,eq}}= \beta_{{\rm eq}} = \beta_{{\rm i}} \left( 
                  \frac{M_{{\rm eq}}}{M_{{\rm H}}} \right)^{1/2} \,,
\end{equation}
where $M_{{\rm eq}} \sim 10^{50}$ g is the horizon mass at
matter--radiation equality. Eq.~(\ref{opbh}) then leads to the
constraint, $\sigma(M_{\odot}) < 0.118$. Due to the exponential
dependence of $\psi(M_{{\rm BH}})$ on $\sigma$, if $\sigma(M_{\odot})$
is reduced below 0.118 by more than a few per-cent, then the present
day density of PBHs becomes negligible. The COBE normalisation gives a
normalisation, on the present horizon scale, of $\sigma(10^{56}
\rm{g}) = 9.5\times 10^{-5}$. Whilst constraints on the abundance of
lighter mass PBHs, due to the consequences of their evaporation (see
e.g. Carr 1996), prevent $\sigma$ from increasing rapidly as $M_{{\rm
H}}$ is decreased. Therefore to produce a non-negligible density of
PBHs with mass $\sim M_{\odot}$, whilst obeying the COBE normalisation
and not over-producing lighter PBHs, the primordial density
perturbation spectrum must have a spike, with finely tuned amplitude,
located at this scale. There are several inflation models which may be
capable of produce such a power spectrum (Ivanov et. al. 1994; Yokoyama
1995; Randall et. al. 1996; Garc\'{\i}a-Bellido et. al. 1996).

\subsection{QCD phase--transition}
The formation of PBHs at the QCD phase-transition was first suggested
by Crawford \& Schramm (1982) At a first--order phase transition the
pressure response of the radiation to compression is reduced due to
the co-existence, in pressure-equilibrium, of a high and a low energy
phase. This leads to a reduction in $\delta_{{\rm c}}$ and PBHs are
formed more easily. The QCD phase-transition occurs at a temperature
$T \sim 100$ MeV, when the horizon mass is $M_{{\rm H}} \sim
M_{\odot}$, PBHs formed during the QCD phase-transition may therefore
naturally have appropriate masses to be viable MACHO candidates.

It is not clear from numerical investigations to date (Jedamzik \&
Niemeyer 1999) if the PBH scaling law (eq.~(\ref{mbh})) holds in this
case; the PBHs formed are typically lighter than the horizon mass and
the spread in the masses appears to be even larger than that found for
those formed during radiation domination.

\section{Microlensing formulae}
In this section we will outline the expressions for the differential
microlensing event rate, for lensing towards the LMC (Griest 1991; De
Rujula et. al. 1991; Alcock et. al. 1996b), including a
non--delta--function mass function. A microlensing event occurs when
the MACHO enters the microlensing `tube', which has radius $u_{{\rm
T}} R_{{\rm E}}$ where $u_{{\rm T}} \approx 1$ is the threshold impact
parameter for which the amplification of the background star is above
the chosen threshold and $R_{{\rm E}}$ is the Einstein radius:
\begin{equation}
R_{{\rm E}}= 2 \left[ \frac{ G M x (1-x)L}{c^2 } \right]^{1/2} \,,
\end{equation}
where $L$ is the distance to the source. Since the distance to the LMC
is much greater than its line of sight depth the sources can all be
assumed to be at the same distance ($\sim 50$ kpc) and the angular
distribution of sources ignored. The MACHO mass is denoted by $M$ and
$x$ is the distance of the MACHO from the observer, in units of $L$.
For any non--delta--function mass function $\psi(M)$\footnote{We will
assume throughout that the MACHO mass function is independent of
position.}, such that the fraction, $f$, of the total mass of the halo
in the form of MACHOs is $f= \int_{0}^{\infty} \psi(M) {\rm d} M$, the
differential event rate (assuming a spherical halo and an isotropic
velocity distribution) is:
\begin{equation}
\label{df}
\frac{{\rm d} \Gamma}{{\rm d} \hat{t}} =  \frac{32 L u_{{\rm T}} }
                 {{\hat{t}}^4
              {v_{{\rm c}}}^2}
                 \int_{0}^{\infty}  \left[ \frac{\psi(M)}{M}  
              \int^{x_{{\rm h}}}_{0} \rho(x) r^{4}_{{\rm E}}
              e^{-Q(x)} \right.
             \times \left.  e^{- \left( v_{{\rm t}}(x) / v_{{\rm c}}
              \right)^2} I_{0}(P(x)) {\rm d} x \right] {\rm d} M \,, 
\end{equation}
where $x_{{\rm h}} \approx 1$ is the extent of the halo, $v_{{\rm
t}}(x)$ is the magnitude of the transverse velocity of the
microlensing tube, $Q(x)= 4 R^{2}_{{\rm E}}(x) u_{{\rm T}}^2 /
(\hat{t}^{2} v_{{\rm c}}^2)$, $ P(x)= 4 R_{{\rm E}}(x) u_{{\rm T}}
v_{{\rm t}}(x)/ (\hat{t} v_{{\rm c}}^2)$ and $I_{0}$ is a Bessel
function. We follow the MACHO collaboration and define $\hat{t}$ as
the time taken to cross the Einstein {\it diameter}. Other
collaborations define the event duration as the Einstein radius
crossing time and their timescales are hence smaller by a factor of 2.

For a standard halo, which consists of a cored isothermal sphere:
\begin{equation}
\rho(R) = \rho_{0} \frac{R_{{\rm c}}^2 + R_{0}^2}{R_{{\rm c}}^2 + R^2} \,,
\end{equation}
where $\rho_{0}= 0.0079 M_{\odot} {\rm pc}^{-3}$ is the local dark
matter density, $R_{{\rm c}} \approx 5$ kpc is the core radius and
$R_{0} \approx 8.5$ kpc is the solar radius, eq.(\ref{df}) becomes
\begin{equation}
\frac{{\rm d} \Gamma}{{\rm d} \hat{t}} = \frac{512 \rho_{0} 
          (R_{{\rm c}}^2 + R_{0}^2) L G^2 u_{{\rm T}} }
             {{\hat{t}}^4 {v_{{\rm c}}}^2 c^4}
                 \int_{0}^{\infty}  \left[ \psi(M) M  
              \int^{x_{{\rm h}}}_{0} \frac{x^2 (1-x)^2}{A + B x + x^2}
           \right.    
             \times \left. e^{\left(-Q(x)\right)} 
              e^{- \left( v_{{\rm t}}(x) / v_{{\rm c}}
              \right)^2} I_{0}(P(x)) {\rm d} x \right] {\rm d} M \,, 
\end{equation}
where $A=(R^2_{{\rm c}}+ R^2_{0})/L^2$, $B=-2(R_{0}/L) \cos{b}
\cos{l}$ and $b=-33^{\circ}$ and $l=280^{\circ}$ are the galactic
latitude and longitude, respectively, of the LMC.

\subsection{Transverse velocities of source and observer}

The transverse velocity of the microlensing tube, $v_{{\rm t}}(x)$, is
often set to zero for simplicity however the motion of the tube
through the halo increases the rate of MACHOs entering it from the
forwards direction and decreases the number leaving it from behind.
This results in an increase in the total event rate, and a decrease in
the average event duration (Griest 1991).  The effect of neglecting
the transverse velocity of the microlensing tube, on the determination
of the MACHO mass function should therefore be investigated. If the
observer has transverse velocity ${\bf v_{{\rm t, o}}}$ and the source
has transverse velocity ${\bf v_{{\rm t, s}}}$ then the transverse
velocity of the microlensing tube (as a function of position along the
tube) is ${\bf v_{{\rm t}}} = (1-x) {\bf v_{{\rm t, o}}} + x {\bf
v_{{\rm t, s}}}$ and it's magnitude, $v_{{\rm t}}= | {\bf v_{{\rm t}}}
|$ is
\begin{equation}
\label{vt}
v_{{\rm t}} = \left[ (1-x)^2 v_{{\rm t, o}}^2 + x^2 v_{{\rm t, s}}^2
            \right.  
       + \left. 2 x (1-x)
    v_{{\rm t, o}}  v_{{\rm t, s}} \cos{\theta} \right]^{1/2} \,,
\end{equation}
where $\theta$ is the angle between $\bf {v_{{\rm t, o}}}$ and
$\bf{v_{{\rm t, s}}}$ (Griest 1991).

The transverse velocities of the LMC and sun, in the rest frame of the
galaxy, in co-ordinates ($v_{{\rm x}}, v_{{\rm y}}, v_{{\rm z}}$)
where $v_{{\rm x}}$ is in the direction of the galactic centre,
$v_{{\rm y}}$ is in the direction of the solar rotation and $v_{{\rm
z}}$ is towards the north galactic pole are ${\bf v_{{\rm LMC}}}=(60,
-155, 144) $ km ${\rm s}^{-1}$ (Jones, Klemola \& Lin 1994) and ${\bf
v_{\odot}}= (9, 231, 6)$ km ${\rm s}^{-1}$, respectively. The
heliocentric position of the centre of the LMC, in galactic
coordinates, is $50.1(0.144, -0.824, -0.548)$ kpc
so that the transverse velocities of the LMC and sun, relative to the
line of sight between them, are
\begin{eqnarray}
{\bf v_{{\rm t, LMC}}}&=& (52, -108, 175) \, {\rm km s^{-1}} \,, \\
{\bf v_{{\rm t, \odot}}}& =& (38, 68, -92) \, {\rm km s^{-1}} \,.
\end{eqnarray}
Inserting these values in eq.(\ref{vt}) gives
\begin{equation}
v_{{\rm t}} =  \left[ (1-x)^2
          \right.  
            + \left. 3.07 x^2 -2.93 x (1-x) \right]^{1/2}
            121 {\rm km s^{-1}} \,.
\end{equation}

\subsection{Halo models}
The standard halo model used above has a number of deficiencies (see
Alcock et. al. 1995 and references therein): the halo may not be
spherical (N body simulations of gravitational collapse produce
axisymmetric or triaxial halos, and several other spiral galaxies
appear to have flattened halos (Sackett et. al. 1994)), the effect of
the galactic disk is neglected leading to an overestimate of the mass
of the halo and the rotation curve of the galaxy may not actually be
exactly flat. The power-law halo models of Evans (1993, 1994) provide
an analytically tractable framework for investigating the effect of
varying the halo model properties on the differential microlensing
rate and hence mass function determination (Alcock et. al. 1995). The
parameters of these models (in addition to the core radius and solar
radius) are: $q$ the axis ratio of the concentric equipotential
spheroids of the halo, $\beta$ which governs the asymptotic behaviour
of the rotation curve $v_{{\rm c}} \sim R^{-\beta}$ and $v_{0}$, the
normalisation velocity which determines the typical MACHO velocities.
The expressions for the differential microlensing rate for these
models can be found in Appendix B of Alcock et. al. (1995).

Other possible halo structures have been considered by various
authors.  De Paolis, Ingorsso \& Jetzer (1996) have investigated the
effect of an anisotropic halo velocity distribution on MACHO mass
determination, using the mass moment method. Markovic \& Sommer-Larsen
(1997) investigated the errors in the determination of the parameters
of a power law mass function which result from assuming a standard
halo model if the MACHOs are actually concentrated towards the
galactic centre, with a velocity dispersion which changes with radius
like that of blue horizontal branch field stars.

It should be noted that analytic descriptions of the halo neglect the
substructure which is observed (Helmi et. al. 1999), and found in
N-body simulations (Moore et. al. 1999). Widrow and Dubinski (1998)
have investigated hypothetical microlensing observations carried out
in a galaxy constructed via a N-body simulation. Whilst they found
that the fraction of lines of sight through the halo which intersect
clumps of matter is small ($\sim 1 \%$), the resulting systematic
errors along these lines of sight are large.

\section{Statistical Method}
The maximum likelihood method has been used to determine the best-fit
parameters for delta--function and power--law mass functions (Alcock
et. al. 1996b; Mao \& Pacz\'{n}yski 1997; Alcock et. al. 1997a;
Markovic \& Sommer-Larsen 1997). We follow the MACHO collaboration and
define the likelihood of a given model as the product of the Poisson
probability of observing $N_{{\rm obs}}$ events when expecting
$N_{{\rm exp}}$ events and the probabilities of finding the observed
durations $\hat{t}_{{\rm j}}$ (where $j=1,....,N_{{\rm obs}}$) from
the theoretical duration distribution, $\mu_{{\rm j}}$, (Alcock
et. al. 1996b; 1997a):
\begin{equation}
\label{likedef}
{\mathcal L} = \exp{\left(-N_{{\rm exp}}\right)} \Pi^{N_{{\rm obs}}}_{j=1}
\mu_{{\rm j}} \,. 
\end{equation}
The expected number of events is given by
\begin{equation}
N_{{\rm exp}} = E \int_{0}^{\infty} \frac{{\rm d} \Gamma}{{\rm d} \hat{t}}
           \,  \epsilon(\hat{t}) \, {\rm d} \hat{t}
\end{equation}
and $\mu_{{\rm j}}$ by
\begin{equation}
\mu_{{\rm j}} = E \, \epsilon(\hat{t}_{{\rm j}}) \, \frac{{\rm d} \Gamma
(\hat{t}_{{\rm j}})}{{\rm d} \hat{t}} \,,
\end{equation}
where $E=1.82 \times 10^{7}$ star years is the exposure,
$\epsilon(\hat{t})$ is the detection efficiency and $\hat{t}$ is the
estimated event duration taking account of blending. We use the same
analytic form for the detection efficiency as Mao \& Paczy\'{n}ski
(1996), but with parameters chosen to give a better fit to the
photometric efficiency, which allows for the effects of blending, of
the MACHO 2-year data:
\begin{eqnarray}
\label{eff}
\epsilon(\hat{t}) = \left\{ \begin{array}{ll}    
               0.3  \exp{\left[-0.394( \ln{t^{\prime}} )^{2.7}\right]}
                
                   & \mbox{if $t^{\prime} >$ 1} \\
                0.3  \exp{\left[-0.281( \ln{t^{\prime}} )^{1.75}\right]}   
                & \mbox{if $t^{\prime} < 1$ } \end{array}.
                                    \right.             
\end{eqnarray}
where $t^{\prime}= \hat{t}/75 {\rm days}$.

\section{Mass function models}
\label{secmfs}
We consider four forms for the MACHO mass function:
\begin{enumerate}
\item Delta--function (DF)
 
A delta--function mass function at $M_{{\rm DF}}$, comprising a
fraction $f$ of the total mass of the halo.

\item Power law--fixed upper cut off (PLF)

A power-law mass function, with exponent $\alpha$, between $M_{{\rm
min}}$ and $M_{{\rm max}}$. We follow Alcock et. al. (1997a) here and
fix $M_{{\rm max}}$ at $12 M_{\odot}$:
\begin{eqnarray}
\label{pl}
\psi (M) = \left\{ \begin{array}{ll}    
               A_{{\rm n}} M^{\alpha}                
                   & \mbox{if $M_{{\rm min}} < M < M_{{\rm max}}$ } \\   
           0     & \mbox{otherwise } \end{array}.
                                    \right.
\end{eqnarray}
where $A_{{\rm n}}$ is a normalisation constant.

\item Power law--variable upper cut off (PLV)

A power-law mass function, as given by Eq.(\ref{pl}), but with the
upper cut-off mass, $M_{{\rm max}}$, allowed to vary.

\item Symmetric power law (SPL)

A symmetric power--law mass function with centre $M_{{\rm c}}$, width
$2 M_{\Delta}$, slope $\alpha$ and normalisation $A_{{\rm n}}$:
\begin{eqnarray}
\label{spl}
\psi (M) = \left\{ \begin{array}{ll}    
               A_{{\rm n}} \left[ 1 - \left(M_{{\rm c}}-M
             / M_{{\rm c}} \right)^{\alpha}   \right]             
                   & \mbox{if $M_{{\rm c}}-M_{\Delta} < M < 
               M_{{\rm c}}$ } \\
                A_{{\rm n}} \left[ 1 - \left(M/ M_{{\rm c}} -M_{{\rm c}}
              \right)^{\alpha}   \right]             
                   & \mbox{if $M_{{\rm c}} < M < M_{{\rm c}} + 
                M_{\Delta}$ } \\
           0     & \mbox{otherwise } \end{array}.
                                    \right.
\end{eqnarray}

\item Primordial black hole (PBH)

The PBH mass fraction derived in Sec.~\ref{defpbh}, which has 2
parameters: the horizon mass at the time the PBHs form, $M_{{\rm H}}$,
and the mass variance, at horizon crossing, on this scale,  $\sigma$. The
fraction, $f$, of the halo mass in MACHOs is identical to $\beta$ the
fraction of the energy density of the universe in PBHs.

\end{enumerate}

\section{Current data}
\label{cur}

In their analysis of the 2-year MACHO collaboration data Alcock
et. al. (1997a) form a 6 event sub-sample by excluding the binary lens
(event 7 in Table~\ref{that}), which may be in the LMC as discussed in
the introduction, and another event (No. 8 in Table~\ref{that}), which
they consider to be the weakest of the 8 events. They argue that this
sub-sample is a conservative estimate of the events resulting from
lenses located in the Milky Way halo.  The effect of microlensing by
other populations (LMC disk and halo, Milky Way disk, spheroid and
bulge) would be more accurately accounted for by including terms
representing their contributions to the expression for the total
differential rate used in Eq.~(\ref{likedef}), as in Alcock
et. al. (2000). Whilst using this more sophisticated method would
increase the uncertainty in, and change the values of, the parameters
of the best fit mass functions it would not change our conclusions
about the ability of the current data to differentiate between mass
function/halo model combinations.

Since we completed our analysis the MACHO project 5.7 year results
have been released. The total number of events is now 13 or 17,
depending on the selection criteria, which corresponds to an optical
depth, and hence halo mass fraction, roughly $50\%$ smaller than that
of the 2 year data. The spread in the timescales of all 13/17 events
is larger than that of the events observed during the first 2 years.
This is at least partly due to the increase in the sensitivity to
longer durations events, which occurs in any survey as the survey
duration increases (Evans \& Kerins 1999). Repeating our analysis
using the new data would change the best fit mass functions, in
particular the values of $f$ found would be reduced by $\sim 50\%$,
but would not change our general conclusions.

We find the maximum likelihood fit, to the 6 event `halo sub--sample',
for each of the mass functions described above for four sample halo
models: the standard halo (SH), the standard halo including the
transverse velocity of the line of sight (SHVT) and 2 power--law halo
models. The power--law halo models used are models B (massive halo
with rising rotation velocity) and C (flattened halo with falling
rotation velocity) from Alcock et. al. (1995) with $\beta= -0.2$
(0.2), $q=1$ (0.78), $R_{{\rm c}}= 5$ kpc (10 kpc), $v_{0}=200$ km
${\rm s}^{-1}$, (210 km ${\rm s}^{-1}$) and $R_{0}$= 8.5 kpc (8.5 kpc)
respectively.  The differential event rate for each of the halo
models, for a delta-function mass function at $M=M_{\odot}$ with
$f=1$, are shown in Fig. 1. To illustrate the uncertainty due to the
small number of events, for the standard halo model, we also find the
best fit mass functions for the full 8 event sample.

The maximum value of the likelihood for each mass function/halo model
combination, relative to that for the delta-function mass function and
a standard halo, are given in Tables~\ref{ml} and~\ref{8ml}, for the 6
and 8 event samples respectively . The best fit mass functions,
described in Sec.~\ref{secmfs} above, are plotted, with the same,
arbitrary, normalisation but different axes scales, in Figs. 2 and 3
for the 6 event sub--sample and in Fig. 4 for all 8 events and a
standard halo. The parameters of the best fit mass functions are given
in Tables~\ref{dfmf} to \ref{pbhmf}. The results for the full 8
events and a standard halo are denoted by `SH8'.

Given the small number of events, maximising the likelihood function
depends mainly on reproducing the observed number of events; for each
mass function halo model combination the best fit to the six event
halo sub sample has $N_{{\rm exp}}$ within 1\% of 6.00.  For the
standard halo (both with and without the transverse velocity of the
line of sight) and power--law halo B the differential rate for a DF
mass function is comparable with the range of the observed timescales,
and hence the DF mass function produces the largest maximum
likelihood. Despite the large differences in the widths of the best
fit mass functions, the differences between the resulting differential
event rates in the region near their peaks, where the event rate is
non--negligible, are very small, especially when the detection
efficiency is included. This can be seen in Fig. 5 where we plot the
differential event rate, with and without the detection efficiency,
for the best fit DF and PBH mass functions for the standard halo. In
the case of power--law halo C, the differential rate for the DF mass
function is narrower than the range of observed timescales so that
broader mass functions are a better fit to the observed events. When
the transverse velocity of the line of sight is included, for the
standard halo, the DF mass function still has the largest maximum
likelihood, however the mean MACHO mass is increased by $\sim 14 \%$.
We can also see that fixing the upper mass cut off, as in Alcock
et. al. (1997a), forces the power--law mass function to fall off more
rapidly than if the upper cut off is allowed to vary. This is because
there are effectively tight limits, due to the absence of long
duration events, on the number of large mass ($M \gg M_{\odot}$)
MACHOs. For the full set of 8 events, and a standard halo model, the
maximum likelihood is attained for mass functions with finite
width. The DF and PBH mass functions have roughly the same maximum
likelihood, with that of each of the power law mass functions (PLF,
PLV, SPL) being roughly 3-4 $\%$ greater.

The differences in maximum likelihood between mass function/halo model
combinations are small and, unsurprisingly given the small number of
events, it is not possible to differentiate between mass functions
using the current data, even if the halo model is fixed.

\section{Monte Carlo Simulations}

In order to assess how many microlensing events will be necessary to
determine whether the MACHO mass function has significant finite
width, we carried out a number of Monte Carlo simulations assuming a
`broad' MACHO mass function and, for different numbers of events,
compared the fit to the data of `broad' and `narrow' mass
functions. To minimise the considerable computing time required for
these simulations we assume a standard halo and neglect the transverse
velocity of the line of sight.  The transverse velocity of the line of
sight should of course be included in the analysis of a set of real
microlensing events, as it leads to a shift in the parameters of the
best fit mass function, however it's inclusion, and the precise form
of the halo models chosen, should not change the general conclusions
of this section.  Furthermore whilst the nature of the halo is not
currently well-known leading, as we saw in Sec.~\ref{cur}, to large
uncertainties in the determination of the MACHO mass function
presumably by the time a large (100+ event) survey is completed our
knowledge of the halo structure will have improved. We take our broad
MACHO mass function to be the best-fit PBH mass function, with
parameters $M_{{\rm H}}=0.70 M_{\odot}$ and $\sigma=0.1153$ as found
by the maximum likelihood analysis of the current data and use the DF
mass function as a 2 parameter `narrow' mass function. Whilst a power
law mass function is perhaps a more realistic `narrow' MACHO mass
function we prefer to compare fairly generic `broad'\footnote{Since
the PBH mass function is close to gaussian it is a reasonable generic
form for a broad mass function} and `narrow' mass functions, with the
same number of free parameters.

We produced 400 simulations each for $N=100,316,1000$ and 3162 events,
for both a perfect detection efficiency ($\epsilon=1$ for all
$\hat{t}$) and that of the first 2 years of the MACHO project, given
by Eq.~(\ref{eff}). The actual efficiency of future long duration
microlensing searches is likely to be somewhere between these two
forms with the peak efficiency, and the duration at which it occurs,
increasing with the search duration (Evans \& Kerins 1999). For each
simulation we find the best fit PBH and DF mass functions by
maximising the likelihood as defined in Eq.~(\ref{likedef}). An
alternative definition of the likelihood function (Markovic \&
Sommer-Larsen 1997) is $\Pi_{i=1}^{N} ({\rm d} \Gamma_{{\rm norm}}
/{\rm d} t)$ where the differential event rate is normalised such that
$\int ({\rm d} \Gamma_{{\rm norm}} /{\rm d} t)=1$. This approach,
however, neglects the information about the normalisation of the
differential event rate which would be obtained in any real
microlensing survey, since a particular number of events will be
observed during a known exposure time.

For each simulation we compare the theoretical event rate
distributions produced by the best fit mass functions with those
`observed' using a modified form of the Kolmogorov-Smirnov (KS)
test. We can not use the standard KS test to compare the theoretical
(PBH and DF) event durations with our Monte Carlo `data', since the
parameters of the input mass functions have been estimated from the
`data'. Instead we use our simulations to compute the probability
distribution of the KS statistic $D_{{\rm KS}}$, the maximum distance
between the theoretical cumulative distribution function and that of
the `data', for the best fit real PBH mass function. We then compare
this distribution with the values of $D_{{\rm KS}}$ of the best fit
`false' DF mass functions. This gives us an indication of the number
of events required to differentiate between mass functions. The
fraction of the simulations passing the KS test at a given confidence
level \footnote{By definition 50\% of the `real' best fit mass
functions are accepted at the 50\% confidence level.}, for both the
PBH and DF MACHO mass functions, is shown in Fig. 6 for each value of
N for both forms of the efficiency. Between 316 and 1000 events should
be sufficient to discern whether the MACHO mass function has
significant finite width.

In Fig. 7 we plot 1 and 2 $\sigma$ contours (which contain 68\% and
95\% of the simulations respectively) of the parameters of the PBH
mass function. The parameters of the input mass function are marked
with a cross In Fig. 8 we plot contours of the parameters ($f$ and
$M_{{\rm DF}}$) of the best fit delta-function mass function and in
Fig. 9 contours of the mean mass and halo fraction of the best fit PBH
mass functions. Fitting a DF mass function when the true mass function
is the PBH mass function leads to a systematic underestimation of the
mean MACHO mass by $\sim 15 \%$. The mean and standard deviation of
the best fit values of $M_{{\rm H}}$ and $\sigma$ obtained are
displayed in Tables \ref{msdc} and \ref{msdc2}, for the MACHO 2-year
and flat efficiencies, respectively.  We find, in agreement with Mao
\& Pacz\'{n}yski (1996) and Markovic \& Sommer-Larsen (1997), that
with $\sim 1000$ events it will be possible to determine the
parameters of the MACHO mass function to a few $\%$, if the halo
structure is known.

\section{Conclusions and future prospects}
The MACHO mass ($M \sim 0.5 M_{\odot}$) and halo fraction ($f \sim 0.5
$) favoured by current microlensing data pose severe problems for
stellar MACHO candidate such as faint stars and white dwarves. It was
previously thought (Yokoyama 1998, Green \& Liddle 1999) that the
relatively broad mass function of primordial black holes was likely to
be inconsistent with the durations of the observed microlensing
events. In Sec. \ref{cur} we found that, using the current data, the
likelihood of the best fit primordial black hole mass is comparable to
that of the best fit delta-function, to $\sim \%$. This then led us to
investigate the number of events necessary to determine, assuming that
the lenses are located in the Milky Way halo and that the halo can
accurately be described by a known analytic form, whether the MACHO
mass function has significant finite width. Approximately 500 events
should be sufficient to answer this question and also determine the
parameters of the mass function to $\sim 5\%$. If the halo model is
not known then the number of events necessary is likely to be
increased by a least an order of magnitude (Markovic \& Sommer-Larsen
1997). The use of a satellite, to make parallax measurements of
microlensing events, would however allow simultaneous determination of
the lens location and, if appropriate, mass function and parameters of
the halo model, with of order 100s events (Markovic 1998). If the
MACHOs are PBHs, then the gravitational waves emitted by PBH-PBH
binaries will allow the MACHO mass distribution to be mapped by the
Laser Interferometer Space Antenna (Nakamura et. al. 1997; Ioka
et. al. 1998; Ioka, Tanaka \& Nakamura 1999).

\acknowledgements 

The author would like to thank Bernard Carr, N. Wyn Evans, Simon
Goodwin, Andrew Gould, Cheongo Han, Andrew Liddle, Anne-Laure
Melchior, Jesper Sommer-Larsen and in particular Martin Hendry and
Draza Markovic for useful comments and/or discussions.

\clearpage

\begin{figure}
\label{figratdfhm}
\plotone{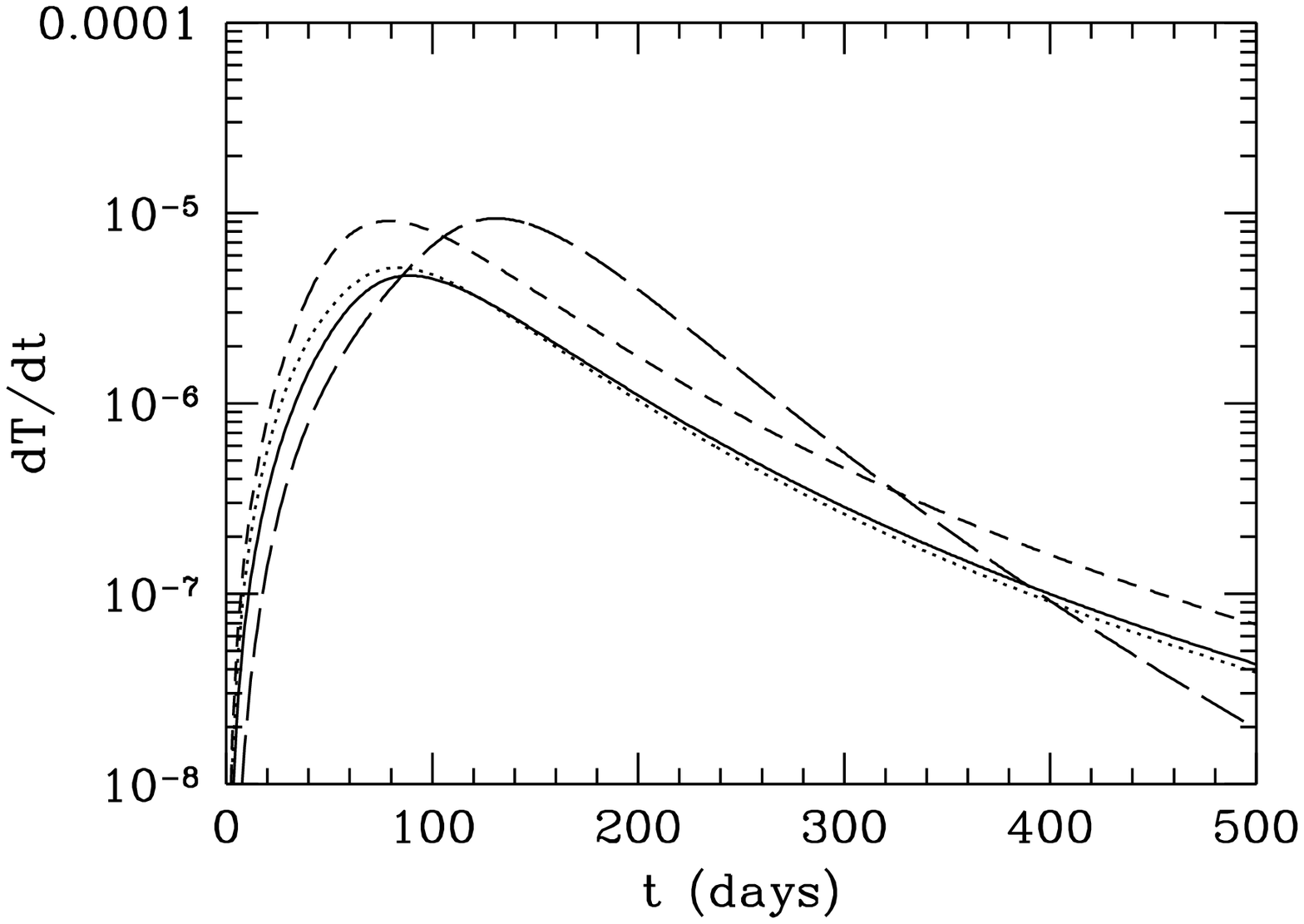}
\caption{The differential rate (in $s^{-2}$) for a delta--function
mass function at $M=M_{\odot}$ with $f=1$ for the standard halo model
(solid line), standard halo including the transverse velocity of the
line of sight (dotted line) and power law halo models B (short dashed
line) and C (long dashed line).}
\end{figure}

\begin{figure}
\plottwo{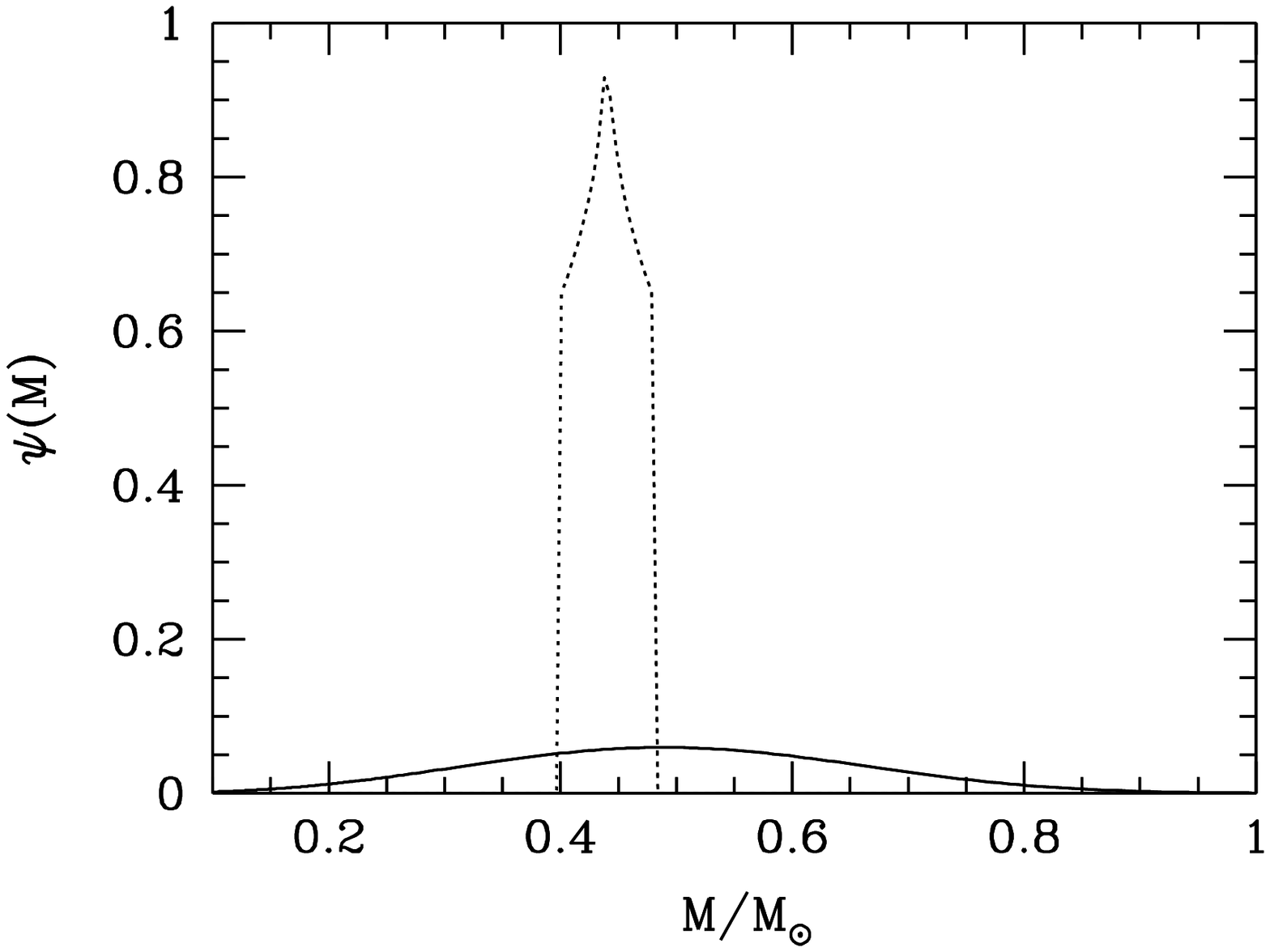}{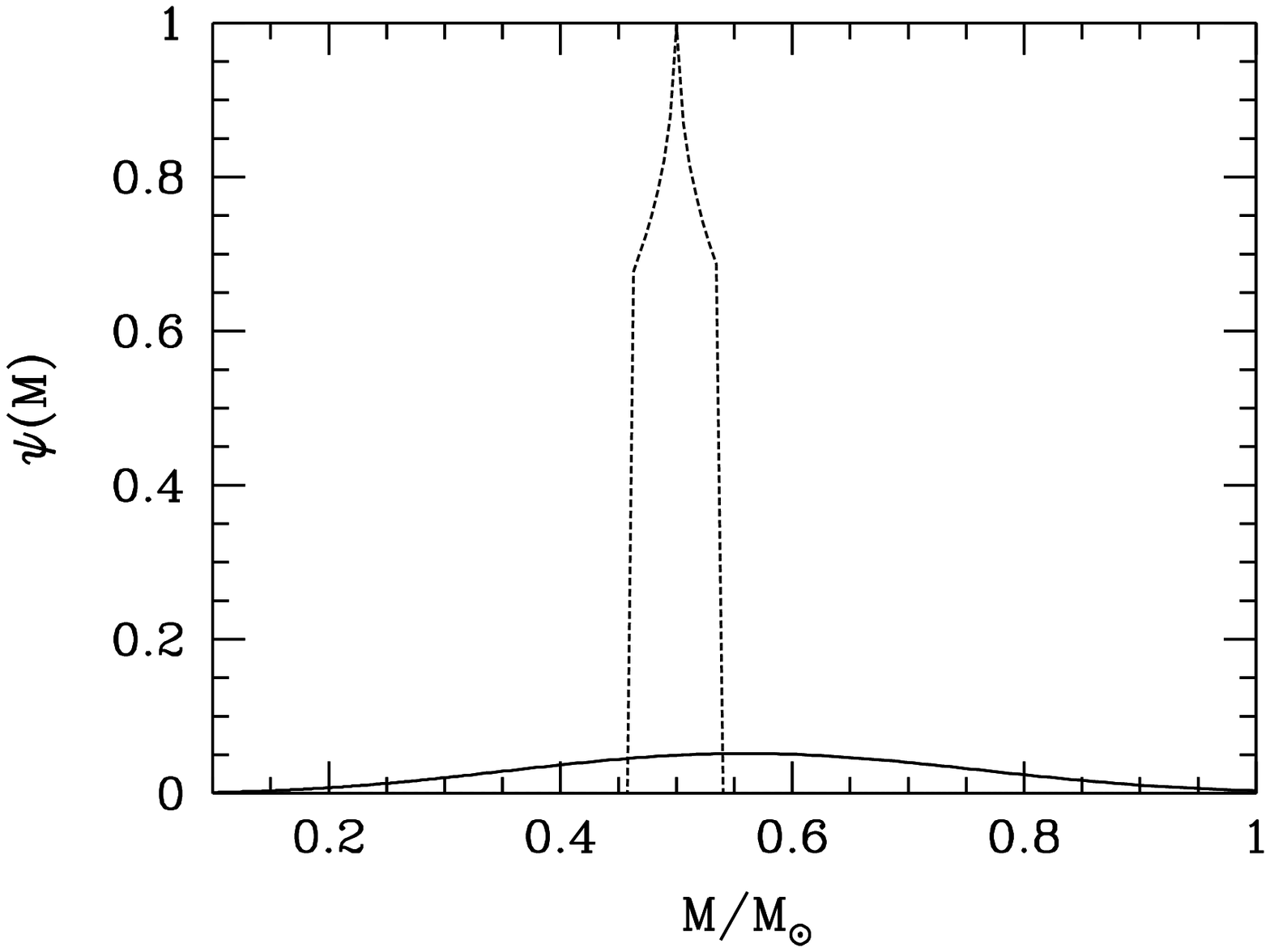}
\label{figmfsh}
\caption{The best fit PBH (solid line) and symmetric power--law (dotted
line) for the
standard halo model (left hand panel) and the standard halo including
the transverse velocity of the line of sight (right hand panel), for
the 6 event sub--sample.}
\end{figure}

\begin{figure}
\plottwo{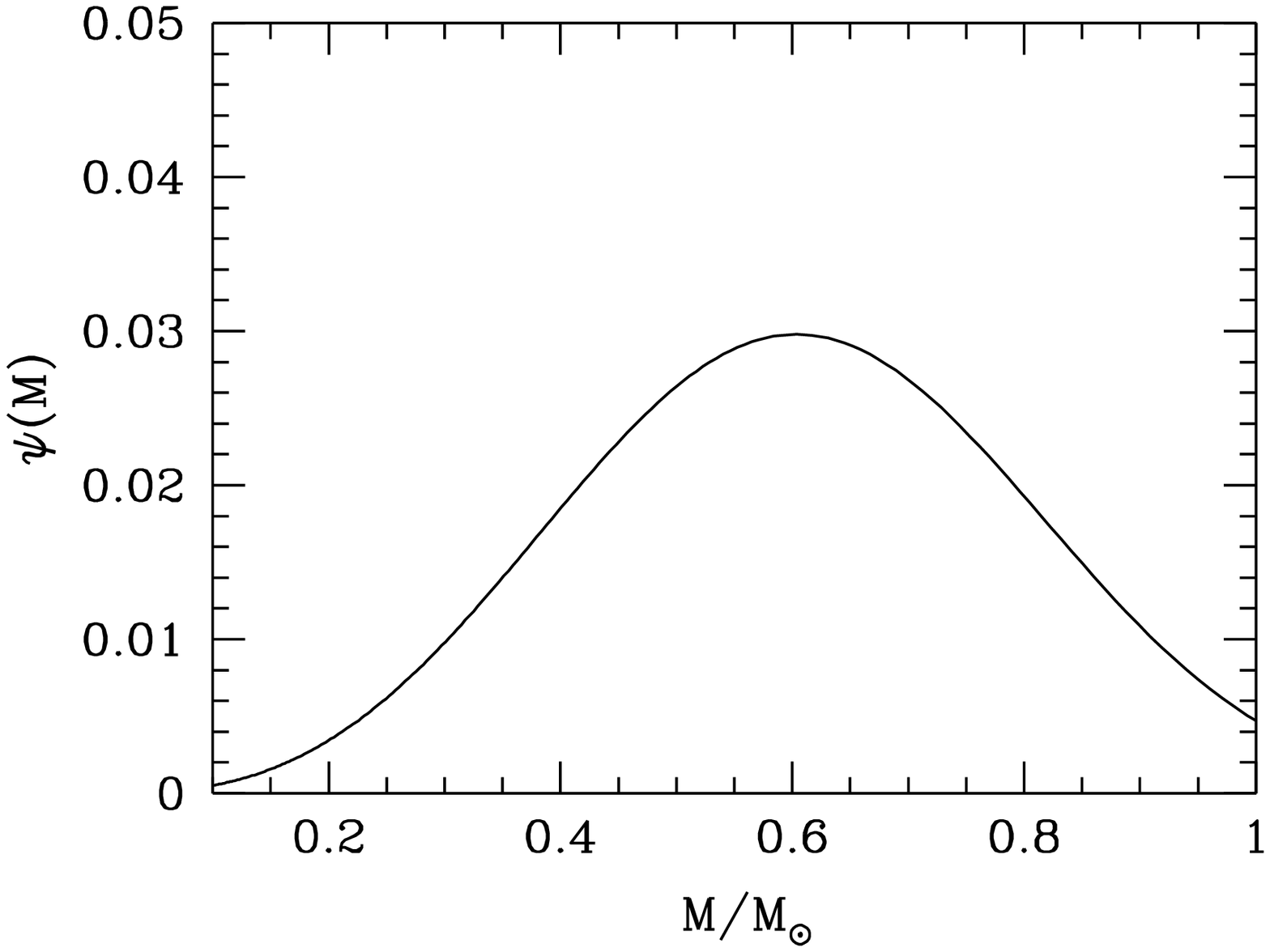}{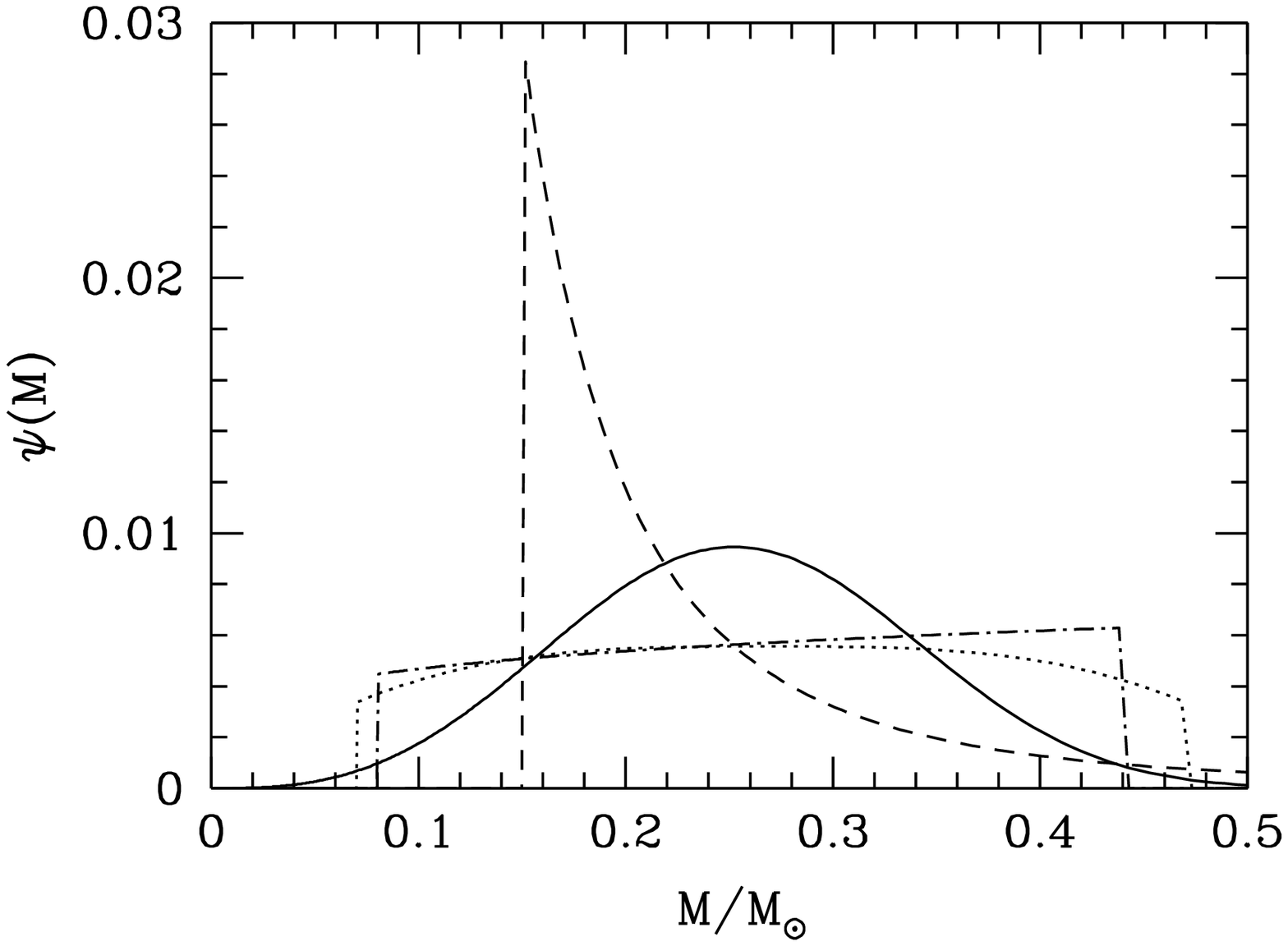}
\label{figmfpl}
\caption{The best fit PBH (solid line), and where appropriate,
symmetric power law (dotted line) power law--fixed upper cut off
(dashed line) and law--free upper cut off (dot dashed) mass functions
for power law halos B (left hand panel) and C (right hand panel), for
the 6 event sub--sample.}
\end{figure}

\begin{figure}
\plotone{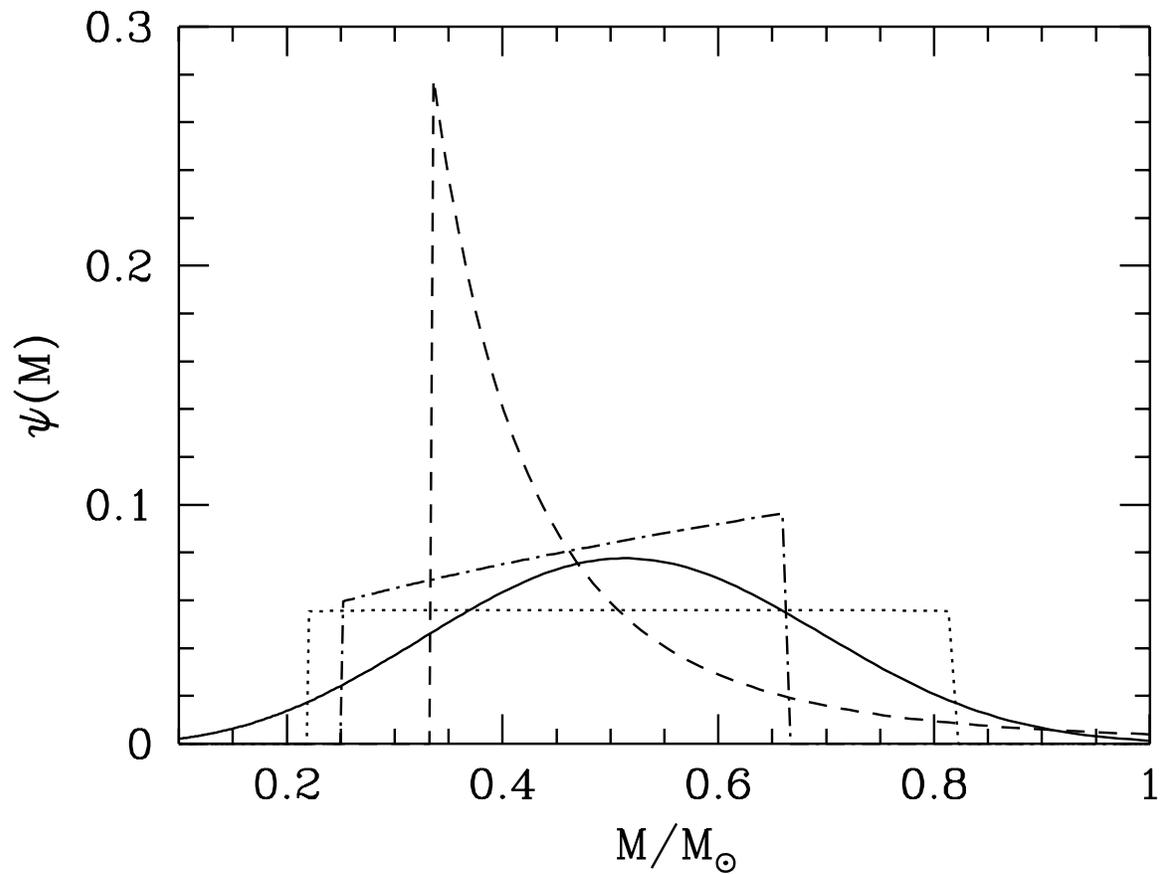}
\label{figmf8}
\caption{The best fit PBH (solid line), symmetric power law (dotted),
power law--fixed upper cut off (dashed line), and power law--free
upper cut off (dot dashed) mass functions, for all 8 events and the
standard halo model.}
\end{figure}

\begin{figure}
\label{figrat}
\plotone{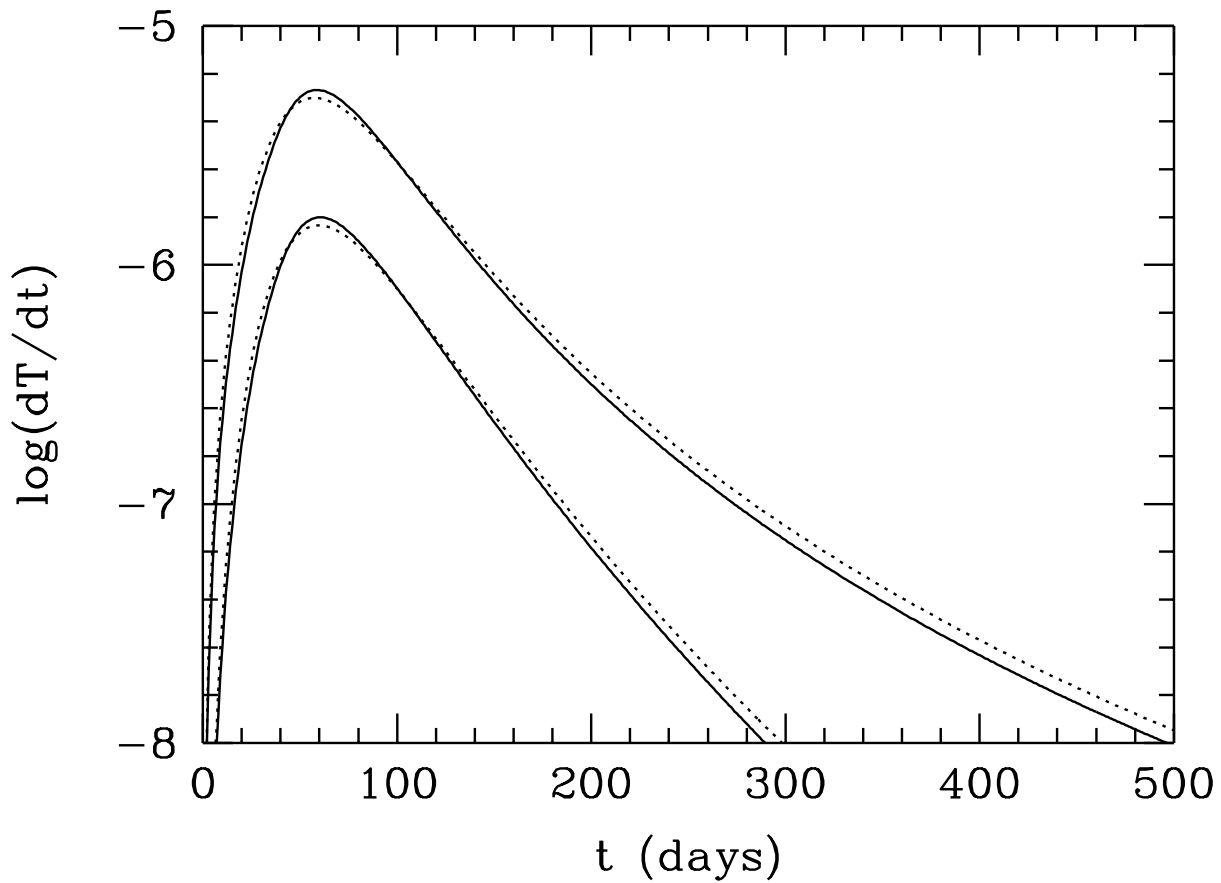}
\caption{The upper and lower curves show the differential rate and the
differential rate multiplied by the experimental efficiency
respectively for the best fit delta-function (solid line) and PBH
(dotted) mass functions, for the 6 event sub--sample and the
standard halo model.}
\end{figure}

\begin{figure}
\plotone{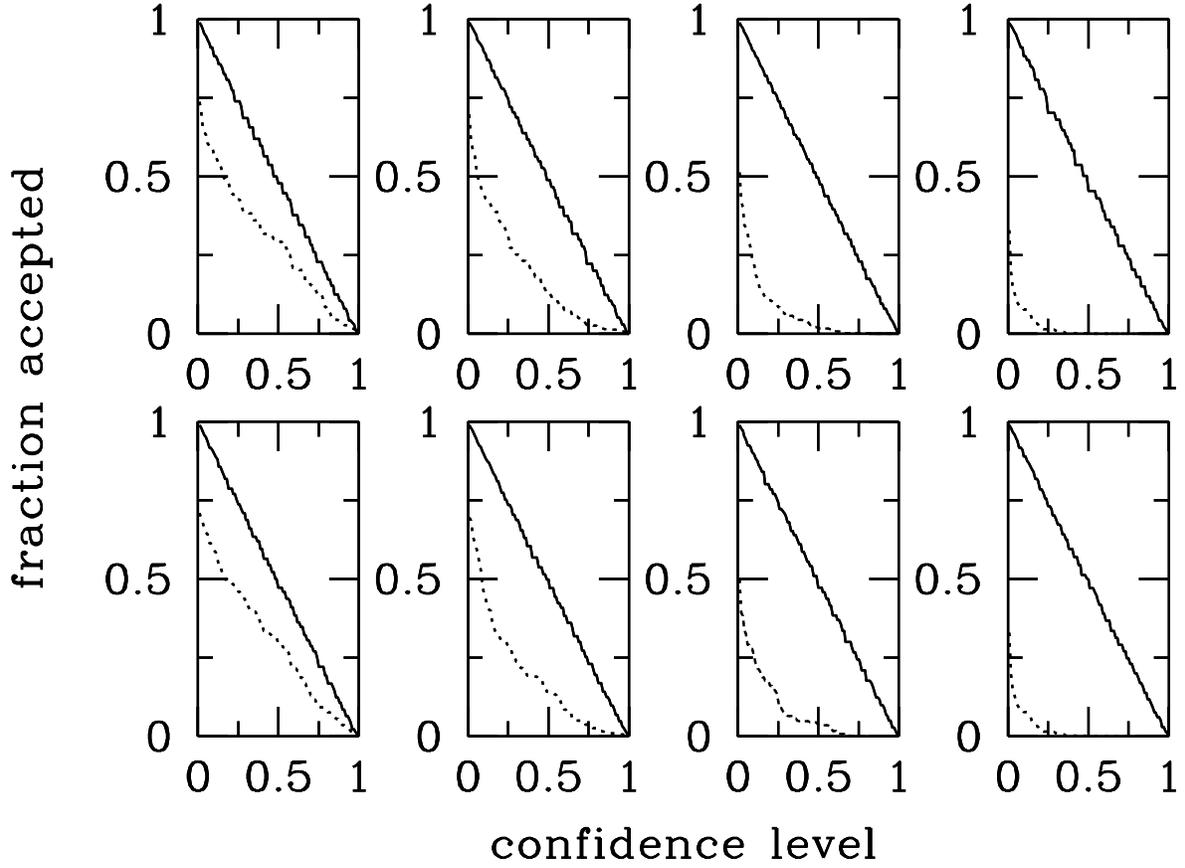}
\label{figprob}
\caption{The fraction of best fit PBH (solid lines) and delta-function
(dotted)mass functions passing the KS test at a given confidence
level, for $N=100, 316, 1000$ and 3162 events from left to right, for
the MACHO 2 year (upper row) and perfect flat (lower) efficiencies.  }
\end{figure}

\begin{figure}
\plottwo{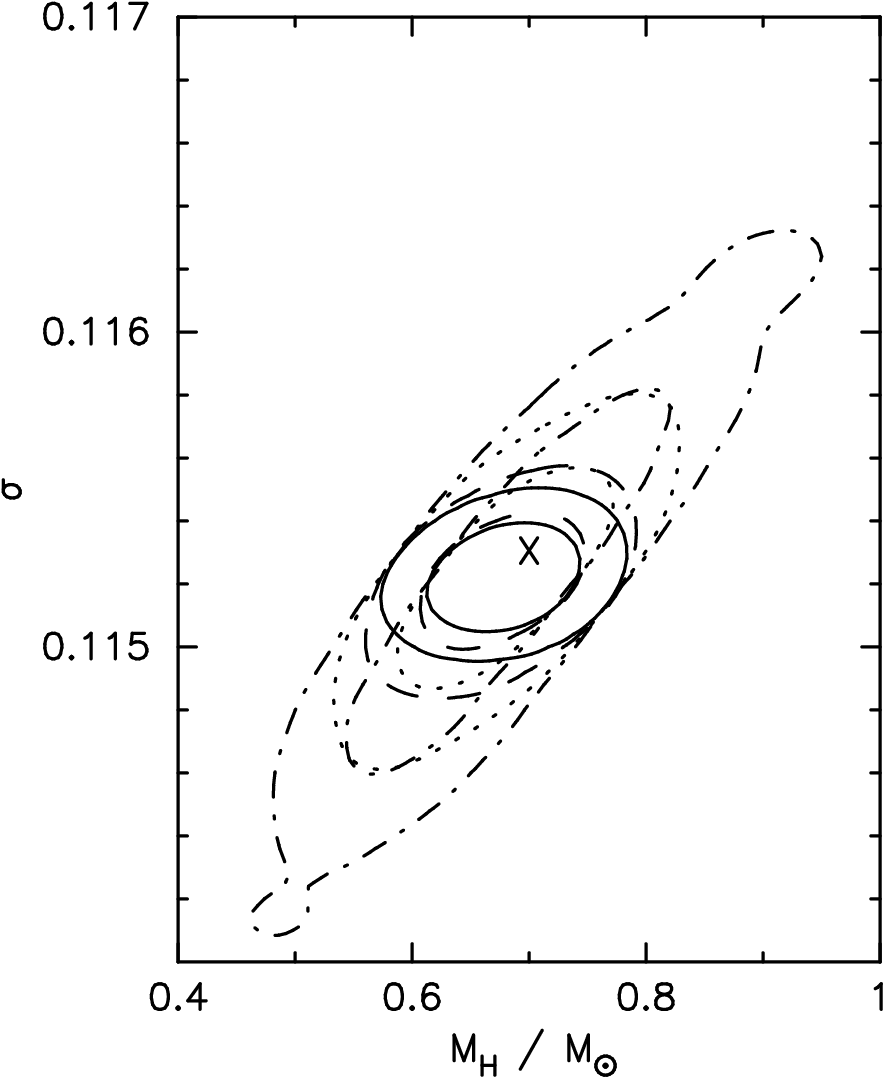}{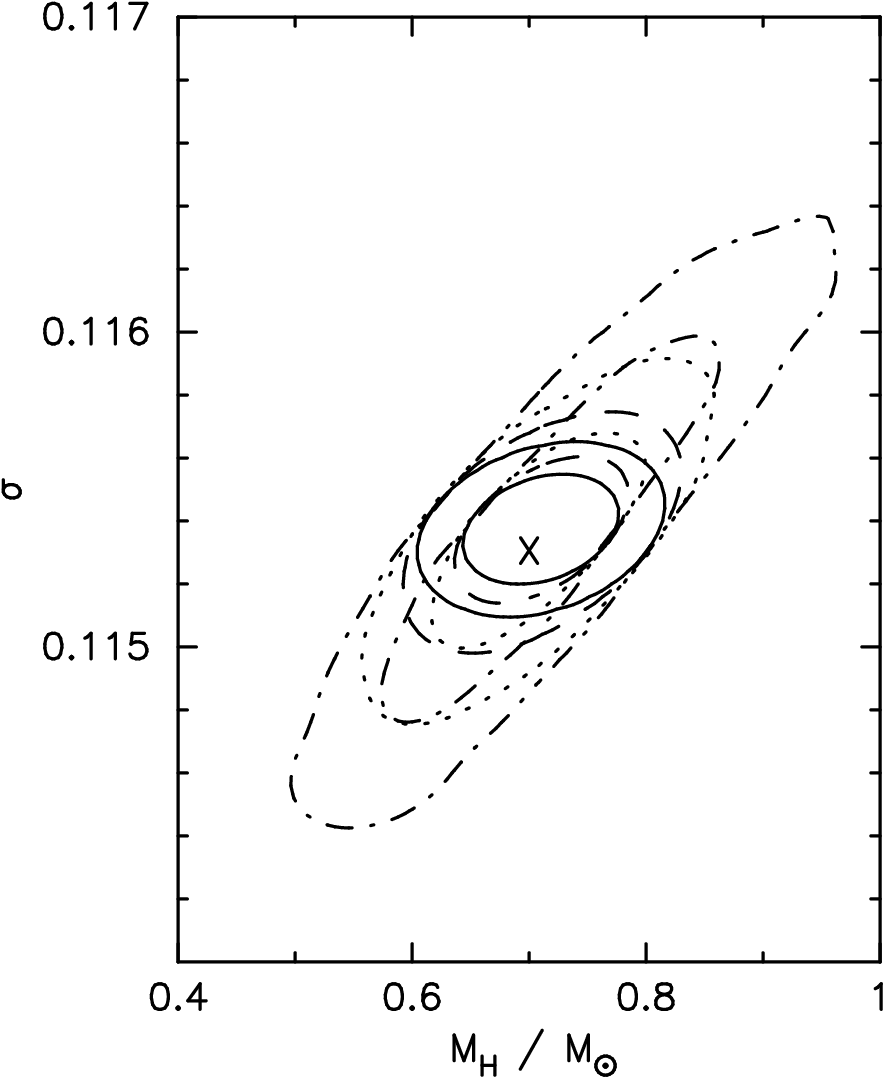}
\label{figcont}
\caption{Contours containing 68 $\%$ and 95 $\%$ of the $N=
100$ (dot-dashed line), $316$ (dotted), $1000$ (dashed) and $3162$
(solid) event best fit PBH MFs for the MACHO 2-year (left panel) and
perfect flat (right panel) efficiencies.}
\end{figure}

\begin{figure}
\plottwo{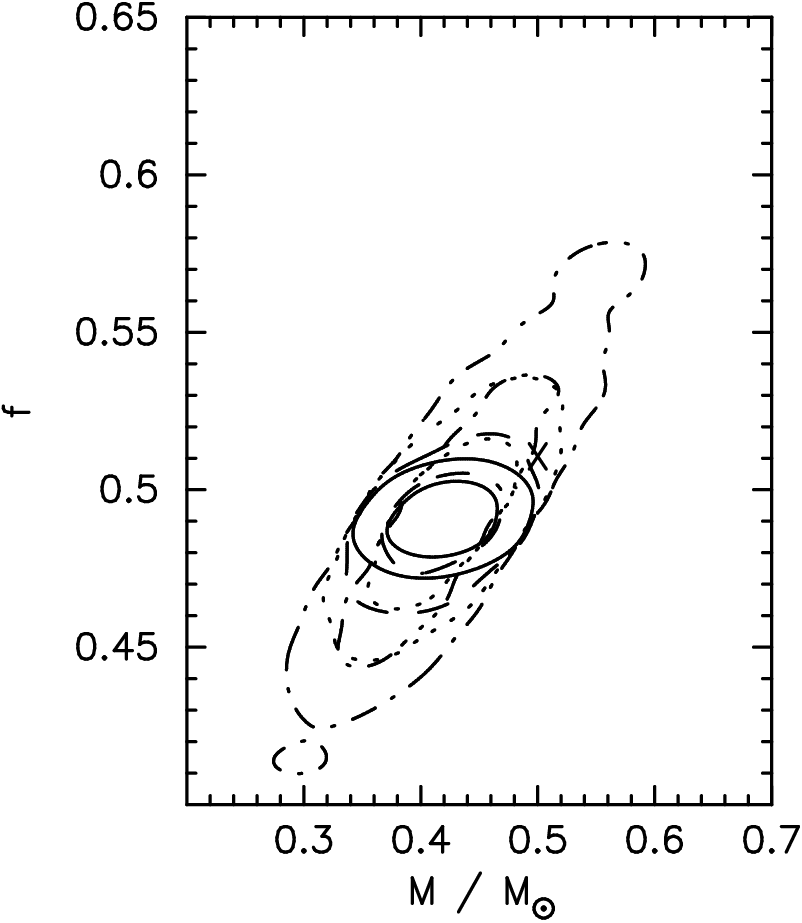}{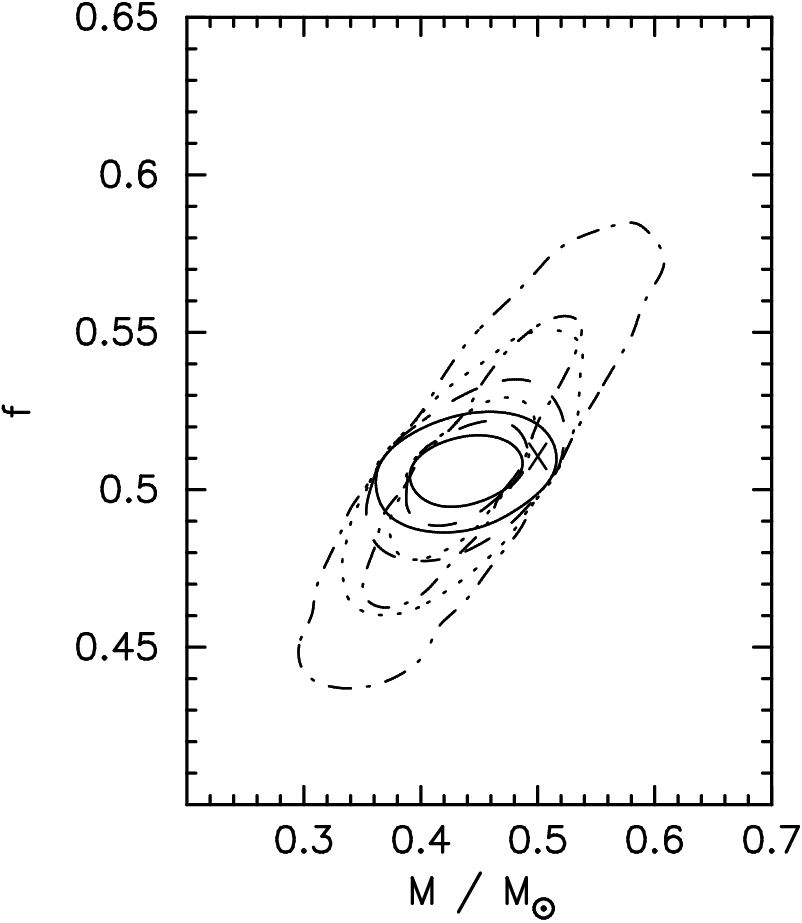}
\label{figdf}
\caption{Contours containing 68 $\%$ and 95 $\%$ of the $N=
100$ (dot-dashed line), $316$ (dotted), $1000$ (dashed) and $3162$
(solid) event best fit DF MFs for the MACHO 2 year (left panel) and
perfect flat (right panel) efficiencies.}
\end{figure}

\begin{figure}
\plottwo{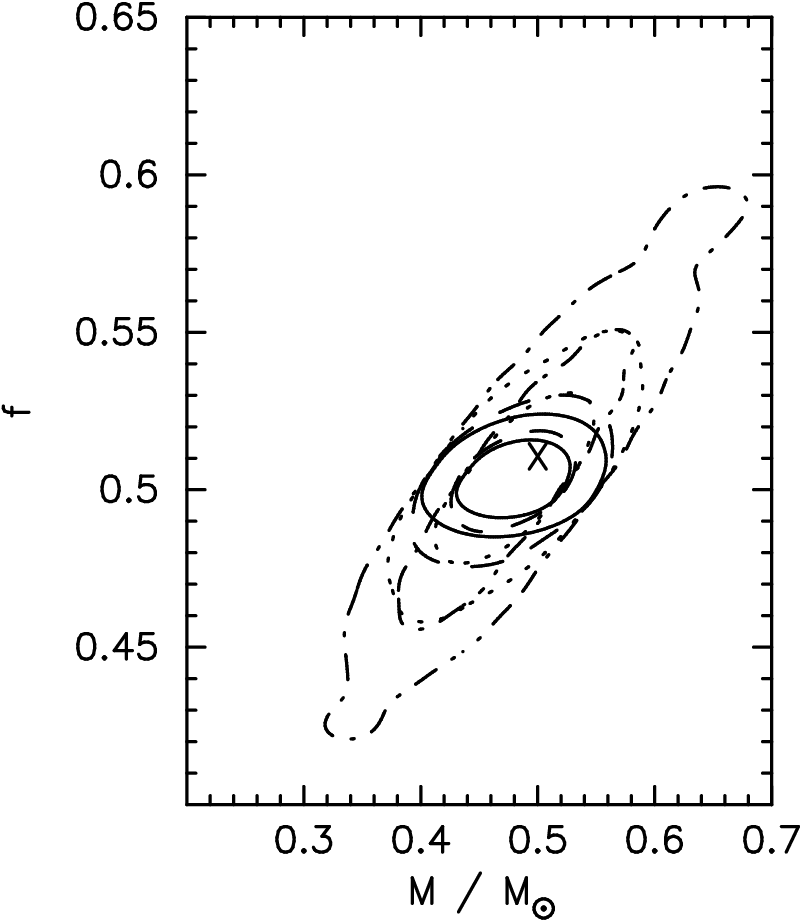}{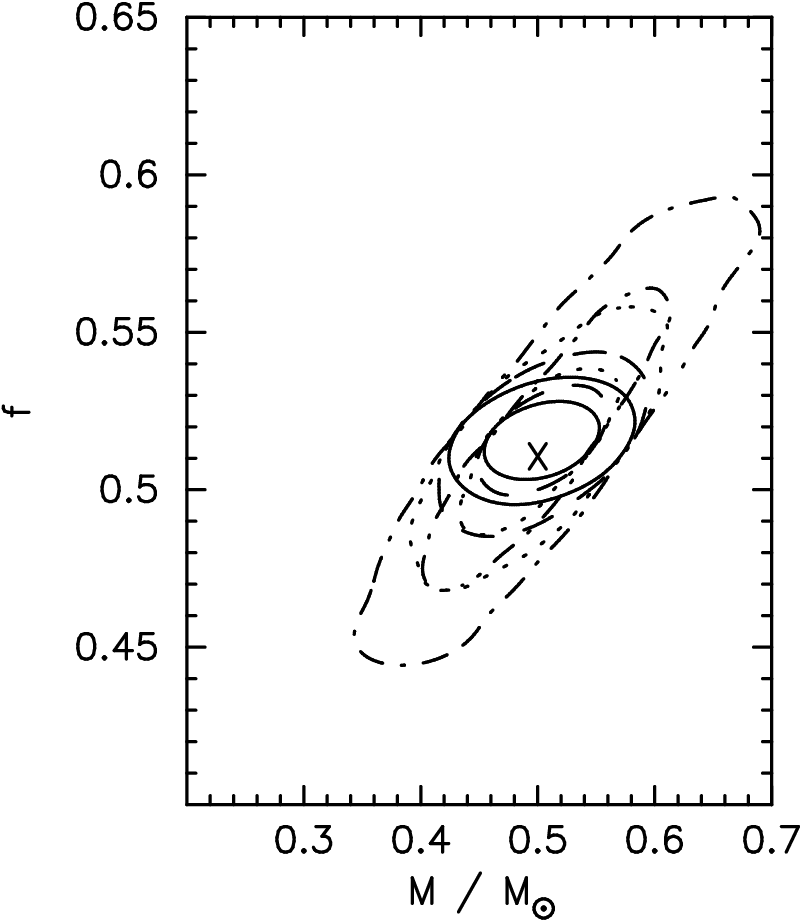}
\label{figcontmf}
\caption{Contours containing 68 $\%$ and 95 $\%$ of the $N=
100$ (dot-dashed line), $316$ (dotted), $1000$ (dashed) and $3162$
(solid) event of the halo fraction and mean MACHO mass for the best
fit PBH MFs for the MACHO 2 year (left panel) and perfect flat (right
panel) efficiencies.}
\end{figure}

\clearpage

\begin{deluxetable}{cc} 
\tablecolumns{2} 
\tablewidth{0pc} 
\tablecaption{The estimated duration, taking into account blending,
of the 8 microlensing events from Alcock et. al (1997a). \label{that} } 
\tablehead{ 
Event & $\hat{t}$ (days)}
\startdata 
1& 38.8 \\ 2 & 52 \\ 3 & 88 \\ 4 & 100 \\ 5 & 131 \\ 6 & 70 \\ 7 & 143 
\\ 8 & 47 \\ 
\enddata 
\end{deluxetable}

\begin{deluxetable}{ccccc} 
\tablecolumns{5} 
\tablewidth{0pc} 
\tablecaption{Relative maximum likelihood of each mass function/halo model
 combination, for the 6 event sub--sample.  \label{ml} } 
\tablehead{ 
Mass function & SH & SHVT & B & C}
\startdata 
DF & 1 & 0.97 & 0.99 & 0.98 \\
PLF& $\rightarrow 1$ & $\rightarrow 0.97$ & 
$\rightarrow 0.99$ & 1.02 \\
PLV & $\rightarrow 1 $ & $\rightarrow 0.97$ & $\rightarrow 0.99$ & 1.12\\
SPL & 1 &0.97  & 1.00 & 1.15 \\
PBH & 0.92  & 0.96 & 0.92 & 1.10 \\
\enddata 
\end{deluxetable}

\begin{deluxetable}{cc} 
\tablecolumns{2} 
\tablewidth{0pc} 
\tablecaption{Relative maximum
likelihood of each mass function model, for all 8 events and a standard halo.
\label{8ml} }
\tablehead{ Mass function & SH8} 
\startdata
DF & 1.00  \\ 
PLF& 1.03 \\
PLV &  1.03 \\ 
SPL & 1.04 \\
PBH & 1.00 \\ 
\enddata 
\end{deluxetable}

\begin{deluxetable}{ccc} 
\tablecolumns{3} 
\tablewidth{0pc} 
\tablecaption{Parameters of best fit delta-function mass functions. 
\label{dfmf} } 
\tablehead{ 
Halo model & $M_{\rm df}/M_{\odot}$& f }
\startdata 
SH & 0.44 & 0.50 \\
SHVT & 0.50 & 0.50 \\
B & 0.53 & 0.30 \\
C & 0.22 & 0.041 \\
SH8 & 0.44 & 0.68 \\
\enddata 
\end{deluxetable}

\begin{deluxetable}{cccccc} 
\tablecaption{Parameters, and halo fraction and mean MACHO mass, of best fit
power law mass function,  with fixed upper cut off $M_{{\rm max}}=
12 M_{\odot}$. \label{plmf}}
\tablecolumns{6} 
\tablewidth{0pc} 
\tablehead{
Halo model & $M_{{\rm min}}/M_{\odot}$ & $\alpha$ & $ A_{{\rm n}}$ & f & 
$\bar{M}/M_{\odot}$ }
\startdata 
SH &$\rightarrow 0.44 $ &$\rightarrow -\infty$  &n/a  & 0.50 & 0.44 \\
SHVT & $\rightarrow 0.50$ &$\rightarrow -\infty$ & n/a   &  0.50 & 0.50 \\
B & $\rightarrow 0.53$ & $\rightarrow -\infty$ & n/a  & 0.31 & 0.53 \\
C &0.15 & -3.2 & 2.7 $\times 10^{70}$ & 0.042 & 0.27 \\
SH8 & 0.34 & -3.9 & $3.0 \times 10^{95}$ &  0.68 & 0.50 \\
\enddata 
\end{deluxetable}

\begin{deluxetable}{ccccccc}
\tablecolumns{7} 
\tablewidth{0pc} 
\tablecaption{Parameters, and halo fraction and mean MACHO mass, of best fit
power law mass function,  with free upper cut off.  \label{plmf2}}
\tablehead{
Halo model & $M_{{\rm min}}/M_{\odot}$ &$M_{{\rm max}}/M_{\odot}$&  $\alpha$ 
& $ A_{{\rm n}}$ & f & $\bar{M}/M_{\odot}$ }
\startdata
SH &$\rightarrow 0.44$ & $\rightarrow 0.44$  & $\rightarrow -\infty$  &n/a 
&0.50 &0.44 \\
SHVT &$\rightarrow 0.50$  &$ \rightarrow 0.50$  & $\rightarrow -\infty$ 
& n/a & 0.51  &0.50  \\
B &$\rightarrow 0.53$  & $\rightarrow 0.53$ & $\rightarrow +\infty$ &n/a  
& 0.30&0.53 \\
C &0.08 & 0.44 &0.2  & $1.7 \times 10^{-41}$ &0.041 & 0.27 \\
SH8 & 0.25 & 0.66 & 0.5 & $2.75 \times 10^{-50}$ & 0.67 & 0.47 \\
\enddata
\end{deluxetable}

\begin{deluxetable}{ccccccc}
\tablecolumns{7} 
\tablewidth{0pc} 
\tablecaption{Parameters, and halo fraction and mean MACHO mass, of best fit
symmetric power law mass function.  \label{plmf3}}
\tablehead{
Halo model & $M_{{\rm c}}/M_{\odot}$ &$M_{{\Delta}}/M_{\odot}$&  $\alpha$ & 
$ A_{{\rm n}}$ & f & $\bar{M}/M_{\odot}$ }
\startdata
SH &0.44 &  0.04  & 0.4  &$ 4.5 \times 10^{-33}$ 
&0.50 &0.44 \\
SHVT & 0.50 &0.04  &0.4 & $4.2 \times 10^{-38}$   & 0.50 & 0.50 \\
B & 0.53  & $\rightarrow 0$ & $\rightarrow 0$ & n/a & 0.30 &0.52  \\
C &0.27 &0.20  &3.1  &$5.9\times 10^{-35}$  &0.042 & 0.27 \\
SH8 & 0.30 & 0.52 & 8.4 & $5.9\times 10^{-34}$ & 0.69 & 0.52 \\ 
\enddata
\end{deluxetable}

\begin{deluxetable}{ccccc} 
\tablecolumns{5} 
\tablewidth{0pc} 
\tablecaption{Parameters, and halo fraction and mean MACHO mass, of best 
fit PBH mass functions. \label{pbhmf} } 
\tablehead{ 
Halo model &  $M_{{\rm H}}/M_{\odot}$& $\sigma(M_{{\rm H}})$ & $f$ &
$\bar{M}/M_{\odot}$ }
\startdata 
SH &  0.70 & 0.1153 & 0.51 & 0.50 \\
SHVT & 0.80 & 0.1155 & 0.51 & 0.57\\
B & 0.87 & 0.1141 & 0.32  & 0.61\\
C & 0.38 & 0.1071 & 0.042 & 0.26\\
SH8 & 0.73 & 0.1164 & 0.70 & 0.52 \\
\enddata 
\end{deluxetable}

\begin{deluxetable}{ccccc} 
\tablecolumns{5} 
\tablewidth{0pc} 
\tablecaption{Mean and standard
deviation (SD) of best fit values of $M_{{\rm H}}$ and $\sigma$
obtained from Monte Carlo simulations, using the 2 year MACHO collaboration
efficiency. \label{msdc} } 
\tablehead{ \colhead{$N_{{\rm e}}$} &
\colhead{ $\bar{M}_{{\rm H}}/M_{\odot}$} & \colhead{SD ($M_{{\rm
H}}/\bar{M_{\rm H}})$}& \colhead{ $\bar{\sigma}$} & 
\colhead{SD ($\sigma/ \bar{\sigma}$)} }
\startdata 
100 & 0.681 & 0.125 & 0.11519 & $3.6 \times 10^{-3}$ \\ 316
& 0.676 & 0.068 & 0.11518& $1.9 \times 10^{-3}$ \\ 1000 & 0.674 &
0.038 & 0.11517 & $1.0\times 10^{-3}$ \\ 
3162 &0.677 &0.024 &0.11518 & $6.9\times 10^{-4}$ \enddata
\end{deluxetable}

\begin{deluxetable}{ccccc} 
\tablecolumns{5} \tablewidth{0pc} 
\tablecaption{Mean and standard
deviation (SD) of best fit values of $M_{{\rm H}}$ and $\sigma$
obtained from Monte Carlo simulations using the perfect flat 
efficiency.
\label{msdc2}}  
\tablehead{ \colhead{$N_{{\rm e}}$} & \colhead{ $\bar{M}_{{\rm
H}}/M_{\odot}$} & \colhead{SD ($M_{{\rm H}}/\bar{M}_{\rm H}$)}& \colhead{
$\bar{\sigma}$} &\colhead{ SD ($\sigma/\bar{\sigma}$)} } 
\startdata 
100 & 0.714& 0.125 &
0.11534 & $3.6 \times 10^{-3}$ \\ 316 & 0.704& 0.065 & 0.11530& $1.8
\times 10^{-3}$ \\ 1000 & 0.708 & 0.041 & 0.11533& $1.1\times 10^{-3}$
\\ 
3162 &0.708 &0.024 &0.11533 &$6.9 \times 10^{-5}$\enddata
\end{deluxetable}

\end{document}